\documentclass[aps,twocolumn,nofootinbib,amsmath,amssymb,pra,reprint,floatfix,superscriptaddress]{revtex4-1}
\usepackage{perpage}
\MakePerPage{footnote}

\usepackage{graphicx}
\usepackage{dcolumn}
\usepackage{color}
\usepackage{bm}
\newcommand{\ket}[1]{\vert #1 \rangle}

\newcommand{\meanvalue}[3]{\langle #1 \vert #2 \vert #3 \rangle}
\newcommand{\ketbra}[2]{\vert #1 \rangle \langle #2 \vert}
\newcommand{\imm}{{\rm i }}

\begin{document}

\title{Solvable model of dissipative dynamics in the deep strong coupling regime}
\author{M. Bina}
\affiliation{Dipartimento di Fisica, Universit\`{a} di Milano, I-20133 Milano, Italy}
\author{G. Romero}
\affiliation{
Departamento de Qu\'imica F\'isica, Universidad del Pa\'is Vasco-Euskal Herriko Unibertsitatea, Apdo. 644, 48080 Bilbao, Spain}
\author{J. Casanova}
\affiliation{
Departamento de Qu\'imica F\'isica, Universidad del Pa\'is Vasco-Euskal Herriko Unibertsitatea, Apdo. 644, 48080 Bilbao, Spain}
\author{J. J. Garc\'ia-Ripoll}
\affiliation{Instituto de F\'isica Fundamental, CSIC, Serrano 113-bis, 28006 Madrid, Spain}
\author{A. Lulli}
\affiliation{Dipartimento di Fisica, Universit\`{a} di Milano, I-20133 Milano, Italy}
\author{F. Casagrande}
\thanks{Deceased}
\affiliation{Dipartimento di Fisica, Universit\`{a} di Milano, I-20133 Milano, Italy}
 \author{E. Solano}
\affiliation{
Departamento de Qu\'imica F\'isica, Universidad del Pa\'is Vasco-Euskal Herriko Unibertsitatea, Apdo. 644, 48080 Bilbao, Spain}
\affiliation{
IKERBASQUE, Basque Foundation for Science, Alameda Urquijo 36, 48011 Bilbao, Spain }

\date{\today}

\begin{abstract}
We describe the dynamics of a qubit interacting with a bosonic mode coupled to a zero-temperature bath in the deep strong coupling (DSC) regime. We provide an analytical solution for this open system dynamics in the off-resonance case of the qubit-mode interaction. Collapses and revivals of parity chain populations and the oscillatory behavior of the mean photon number are predicted. At the same time, photon number wave packets, propagating back and forth along parity chains, become incoherently mixed. Finally, we investigate numerically  the effect of detuning on the validity of the analytical solution. 
\end{abstract}

\maketitle

\section{Introduction}
The quantum Rabi model~\cite{Rabi36,Braak11}, describing the interaction of a two-level system and a quantized single mode beyond the rotating-wave approximation (RWA), has found an experimental playground with the advent of novel technologies. On one hand, the development of solid-state semiconductors~\cite{Gunter09} and, on the other hand, the impressive progress on superconducting circuits~\cite{Abdumalikov08,Bourassa09,Niemczyk10,Pol10}, have produced the largest light-matter coupling ever observed. In this sense, the ultrastrong coupling (USC) regime happens when the coupling strength $g$ is comparable to appreciable fractions of the oscillator frequency~$\omega$, $0.1 \lesssim~g/\omega \lesssim~1$. Moreover, it is expected that these architectures can reach soon the deep strong coupling (DSC) regime, where $g/\omega \gtrsim~1$. These regimes are unattainable for the usual experiments in quantum optics, be in trapped ions~\cite{Leibfried03} or cavity QED setups~\cite{Raimond01,Walther06}. In absence of dissipation, the DSC regime predicts the appearance of a different kind of collapses and revivals in the qubit level statistics, which are explained as photon number wave packets propagating back and forth along two independent parity chains~\cite{Casanova10}. From this point of view, it would be of fundamental interest to study these key features, and possible analytical solutions, in presence of dissipation~\cite{SDOAL}. Previous works studied related issues, but considering cases between the perturbative JC model with $g/\omega \lesssim 0.1$ and the USC regime~\cite{Werlang08,Milburn10,Dodonov10}.

In this work, we study the DSC regime of the quantum Rabi model in presence of mode dissipation. In Sec.~II, we present our model and show that a zero-temperature Markovian bath drives the system to an incoherent mixture of two parity chains of the Hilbert space. In order to explore the solvability of this novel quantum open system, we focus on the case of a slow qubit: $\omega_0 \ll \{ \omega, g \}$, where $\omega_0$ is the qubit transition frequency. This allows us to explain the persistence of collapses and revivals of the parity chain probabilities and the behavior of the mean photon number in the presence of dissipation. In Sec.~III, we show the main results of our analytical approach. For the case of near-resonance, in Sec.~IV, we provide a detailed numerical analysis of the asymptotic behavior of probability amplitudes associated to states belonging to both parity chains and the photon statistics. Finally, in Sec.~V, we present our concluding remarks.

\section{The analytical model}
We consider a general system composed of one qubit coupled to a single bosonic mode, as described by the quantum Rabi Hamiltonian
\begin{equation}\label{Hamiltonian}
\hat{H}=\hbar\omega\hat{a}^\dag\hat{a}+\frac{\hbar}{2}\omega_{0} \hat{\sigma}_{z}+\hbar g(\hat{\sigma}+\hat{\sigma}^+)(\hat{a}+\hat{a}^\dag),
\end{equation}
where $\hat{\sigma}_{z}$ is a Pauli operator, $\hat{\sigma}$ and $\hat{\sigma}^\dag$ are the lowering and raising qubit operators, $\hat{a}$ and $\hat{a}^\dag$ are the annihilation and creation mode operators, while $\ket{e}$ and $\ket{g}$ are the upper and lower qubit states, correspondingly.

It is well known that the quantum Rabi model in regimes where the RWA can be applied, that is the Jaynes-Cummings model~\cite{JaynesCummings}, is a solvable dynamics. Recently, there has been a renewed interest in analytical and numerical approximations of the quantum Rabi model beyond RWA, that is, in the USC and DSC regimes~\cite{Irish05,Irish07,Ciuti07,Hausinger08,Ashhab10,Hausinger10,Hwang10,Chen10a,Chen10b,Hausinger11,Casanova10}.  
Here, we consider dissipative effects to the previous studied cases. More precisely, we add a reservoir acting only on the bosonic mode subsystem, and neglect dissipation and decoherence of the slow qubit. The dissipative channel will be described by a thermal bath at zero temperature under the Born-Markov approximation. This model holds for the slow qubit approximation and we will prove that this quantum open system still possesses analytical solutions.

The time evolution of the above described system obeys the following master equation (ME)
\begin{equation}\label{ME}
\dot{\hat{\rho}}=-\frac{\imm}{\hbar}[\hat{H},\hat{\rho}]+\hat{\mathcal{L}}_f[\hat{\rho}],
\end{equation}
where $\hat{\mathcal{L}}_f[\hat{\rho}]$ is the standard Liouville superoperator
\begin{equation}\label{Liouville}
\hat{\mathcal{L}}_f[\hat{\rho}]=\frac{\kappa}{2}(2\hat{a}\hat{\rho}\hat{a}^\dag
-\hat{a}^\dag\hat{a}\hat{\rho}-\hat{\rho}\hat{a}^\dag\hat{a})
\end{equation}
and $\kappa$ is a mode decay rate. We move into an interaction picture, rewriting Eq.~(\ref{Hamiltonian}) as $\hat{H}=\hat{H}_0+\hat{H}_1$,
\begin{align}
\hat{H}_0&=\hbar\omega_0\left ( \frac{\hat{\sigma}_z}{2}+\hat{a}^\dag\hat{a}\right)\\
\hat{H}_1&=\hbar\Delta\hat{a}^\dag\hat{a}+\hbar g(\hat{\sigma}+\hat{\sigma}^+)(\hat{a}+\hat{a}^\dag) ,
\end{align}
where we have introduced the detuning parameter $\Delta=\omega-\omega_0$, with $0\leq\omega_0\leq\omega$. By means of the unitary transformation
$\hat{U}(t)=\exp(\imm\hat{H}_0 t/\hbar)$, we find that the structure of Eq.~(\ref{ME}) holds for the density operator $\hat{\rho}_I=
\hat{U}\hat{\rho}\hat{U}^\dag$ with $\hat{H}$ replaced by the interaction Hamiltonian
\begin{equation}\label{H_Int_pict}\begin{split}
\hat{\mathcal{H}}_I&=\hat{U}\hat{H}_1\hat{U}^\dag\\
&=\hbar\Delta\hat{a}^\dag\hat{a}+\hbar g({\rm e}^{2\imm\omega_0t}\hat{\sigma}^+\hat{a}^\dag+\hat{\sigma}^+\hat{a}+\mathrm{H.c.}).
\end{split}\end{equation}

We consider now the qubit rotated basis, $\ket{\pm}=(\ket{g}\pm\ket{e})/\sqrt{2}$, and rewrite the total system density operator $\hat{\rho}_I(t)$, obtaining the four operators
$\hat{\rho}_{\alpha\beta}(t)=\meanvalue{\alpha}{\hat{\rho}_I(t)}{\beta}\,(\alpha,\beta=\pm)$, describing the bosonic mode subsystem. In this way, we obtain the following set of coupled differential equations
\begin{align}\label{SetRho}
\nonumber\begin{split}
\dot{\hat{\rho}}_{++}=&-\imm\Delta[\hat{a}_t^\dag\hat{a}_t,\hat{\rho}_{++}]-\imm g \cos(\omega_0t)[\hat{a}_t+\hat{a}_t^\dag,\hat{\rho}_{++}]\\
&-g\sin(\omega_0t)[(\hat{a}_t+\hat{a}_t^\dag)\hat{\rho}_{-+}-\hat{\rho}_{+-}(\hat{a}_t+\hat{a}_t^\dag)]+\hat{\mathcal{L}}\hat{\rho}_{++}
\end{split}\\
\nonumber\begin{split}
\dot{\hat{\rho}}_{--}=&-\imm\Delta[\hat{a}_t^\dag\hat{a}_t,\hat{\rho}_{--}]+\imm g \cos(\omega_0t)[\hat{a}_t+\hat{a}_t^\dag,\hat{\rho}_{--}]\\
&+g\sin(\omega_0t)[(\hat{a}_t+\hat{a}_t^\dag)\hat{\rho}_{+-}-\hat{\rho}_{-+}(\hat{a}_t+\hat{a}_t^\dag)]+\hat{\mathcal{L}}\hat{\rho}_{--}
\end{split}\\
\nonumber\begin{split}
\dot{\hat{\rho}}_{+-}=&-\imm\Delta[\hat{a}_t^\dag\hat{a}_t,\hat{\rho}_{+-}]-\imm g \cos(\omega_0t)\{\hat{a}_t+\hat{a}_t^\dag,\hat{\rho}_{+-}\}\\
&-g\sin(\omega_0t)[(\hat{a}_t+\hat{a}_t^\dag)\hat{\rho}_{--}+\hat{\rho}_{++}(\hat{a}_t+\hat{a}_t^\dag)]+\hat{\mathcal{L}}\hat{\rho}_{+-}
\end{split}\\
\begin{split}
\dot{\hat{\rho}}_{-+}=&-\imm\Delta[\hat{a}_t^\dag\hat{a}_t,\hat{\rho}_{-+}]+\imm g \cos(\omega_0t)\{\hat{a}_t+\hat{a}_t^\dag,\hat{\rho}_{-+}\}\\
&+g\sin(\omega_0t)[(\hat{a}_t+\hat{a}_t^\dag)\hat{\rho}_{++}+\hat{\rho}_{--}(\hat{a}_t+\hat{a}_t^\dag)]+\hat{\mathcal{L}}\hat{\rho}_{-+}
\end{split}
\end{align}
where $[ , ]$ and $\{$ , $\}$ denote the standard commutator and anti-commutator symbols, while $\hat{a}_t\equiv{\rm e}^{-\imm\omega_0 t}\hat{a}$.

We will study how collapses and revivals behaves in presence of dissipative mechanisms. For this reason, instead of studying the asymptotic character of the previous equation, we will develop an analytical method that works in the case of the slow qubit limit, $\omega_0 \ll \{ g, \Delta \}$. We will consider evolution times under the condition $\omega_0 t\ll 1$, but long enough to permite several collapses and revivals, $\omega t>1$. In this case, we can approximate $\cos(\omega_0 t)\approx 1$ and $\sin(\omega_0 t)\approx 0$. 

In order to solve Eq.~(\ref{SetRho}), we consider the characteristic function associated to the Wigner function, using the four continuous and square-integrable functions $\chi_{\pm\pm}(\alpha,t)\equiv{\rm Tr}[\hat{\rho}_{\pm\pm}(t)\hat{D}(\alpha)]$, where $\hat{D}(\alpha)={\rm e}^{\alpha\hat{a}^\dag-\alpha^*\hat{a}}$ is the displacement operator. Here, we present the analytical solution in the case where the qubit is prepared in its ground state and the mode in the vacuum state, $\hat{\rho}_I(0)=\ketbra{g}{g}\otimes\ketbra{0}{0}$. In this manner, we have
\begin{subequations}\label{char_evolution}
\begin{align}
\chi_{\pm\pm}(\alpha,t)&=\frac{1}{2}{\rm exp}\Big \{-\frac{|\alpha|^2}{2}\mp\beta(t)\alpha^*\pm\beta^*(t)\alpha\Big \},\\
\chi_{\pm\mp}(\alpha,t)&=\frac{1}{2}F(t){\rm exp}\Big \{-\frac{|\alpha|^2}{2}\mp\beta(t)\alpha^*\mp\beta^*(t)\alpha\Big \}.
\end{align}
\end{subequations}
The corresponding operators $\hat{\rho}_{\pm,\pm}(t)$ take the
form
\begin{subequations}\label{rho_field_evolution}
\begin{align}
\hat{\rho}_{\pm\pm}(t)&=\frac{1}{2}\ketbra{\pm\beta(t)}{\pm\beta(t)}\\
\hat{\rho}_{\pm\mp}(t)&=\frac{1}{2}\frac{F(t)}{{\rm e}^{-2|\beta(t)|^2}}\ketbra{\pm\beta(t)}{\mp\beta(t)},
\end{align}
\end{subequations}
where the field coherent states amplitude $\beta(t)$ and the decoherence function $F(t)$ are defined as
\begin{align}
\beta(t) & \equiv\frac{\imm g}{z}\Big( {\rm e}^{-z t}-1\Big ),\\
F(t) & \equiv{\rm e}^{ -\frac{2 g^2}{|z|^2}\left [ \kappa t+\frac{2}{g}\Im{\rm m}(z^*\beta(t)) \right ]},
\end{align}
with the complex variable $z=\kappa/2+\imm\Delta$. Once obtained the time evolution of the whole system density operator $\hat{\rho}_I(t)$, we derive the quantities of interest such as the probability of the system to be in one of the states $\ket{g n}$ or $\ket{e n}$, that is
\begin{equation}
P_{g/e,n}(t)=\frac{1}{2}{\rm
e}^{-|\beta(t)|^2}\frac{|\beta(t)|^{2n}}{n!}\Big [ 1 \pm (-1)^n
F(t){\rm e}^{2|\beta(t)|^2} \Big ],
\end{equation}
where the sign $+$($-$) holds for $g$($e$). In addition, we compute the purity $\mu(t)=Tr[\rho_I^2(t)]$ of the whole system
\begin{equation}\label{purity}
\mu(t)=\frac{1}{2}\left [ 1+\frac{F^2(t)}{{\rm
e}^{-4|\beta(t)|^2}}\right ].
\end{equation}
As regards the qubit and bosonic mode subsystems, we derive the expressions for the qubit level populations $P_{g/e}(t)$ and the photon number distribution $P_n(t)$
\begin{align}
P_{g/e}(t)&=\frac{1}{2}\left [ 1\pm F(t)\right ],\\
P_n(t)&={\rm e}^{-|\beta(t)|^2}\frac{|\beta(t)|^{2n}}{n!},
\end{align}
where $P_n(t)$ is a Poissonian distribution with a mean photon number $\langle N(t) \rangle=|\beta(t)|^2$.
\begin{figure}
\includegraphics[width=0.35\textwidth]{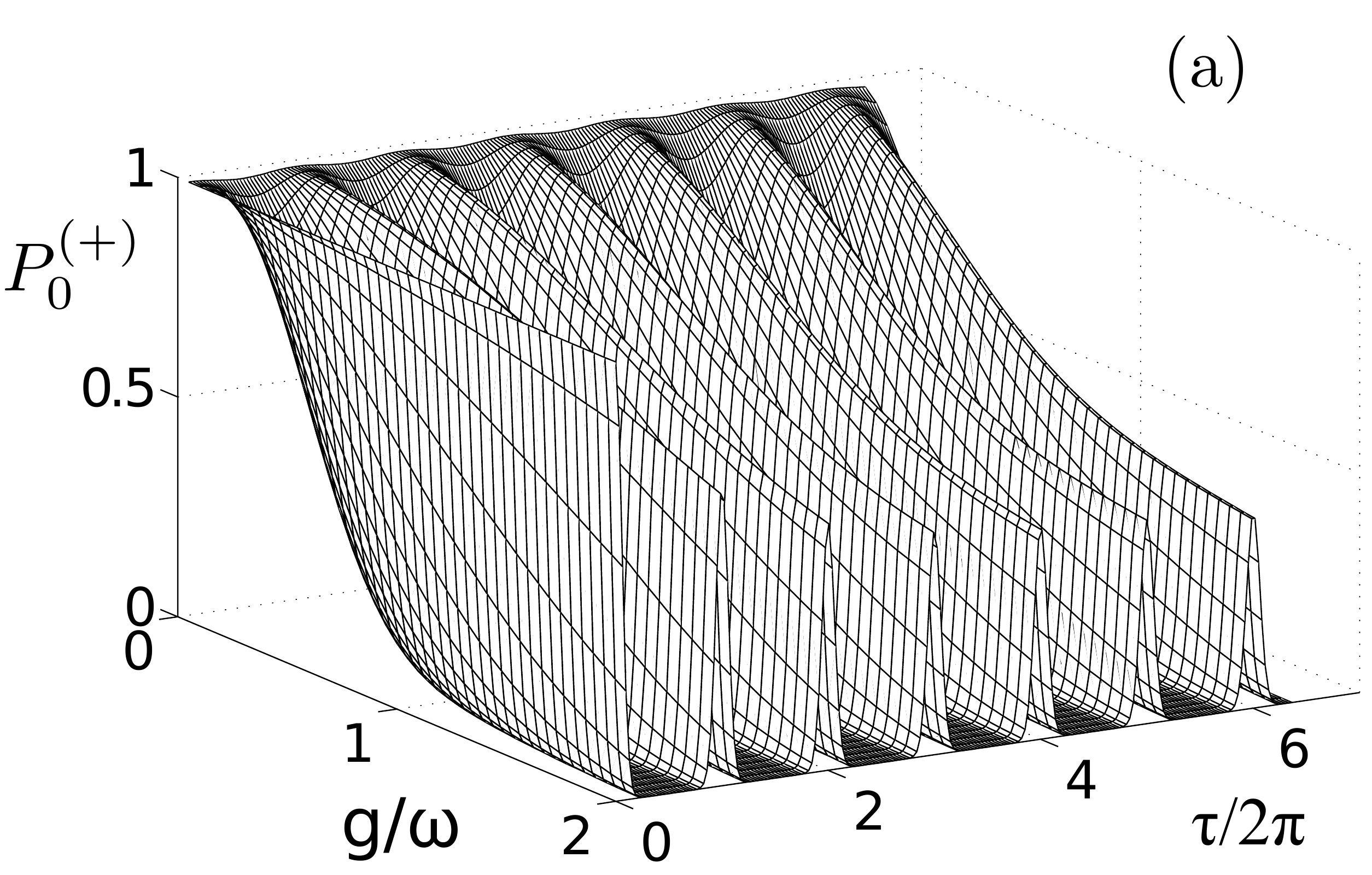}\\
\includegraphics[width=0.35\textwidth]{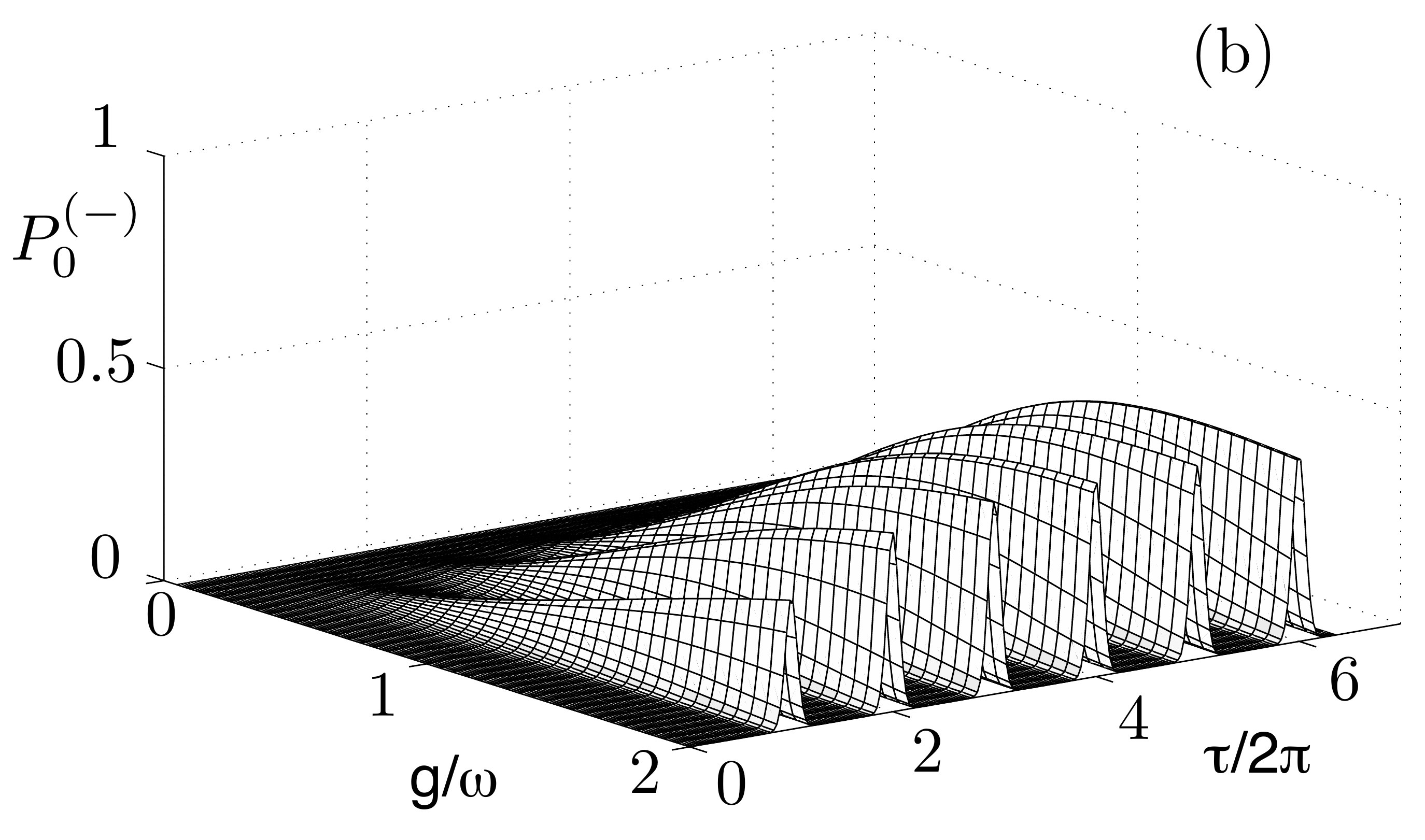}\\
\includegraphics[width=0.35\textwidth]{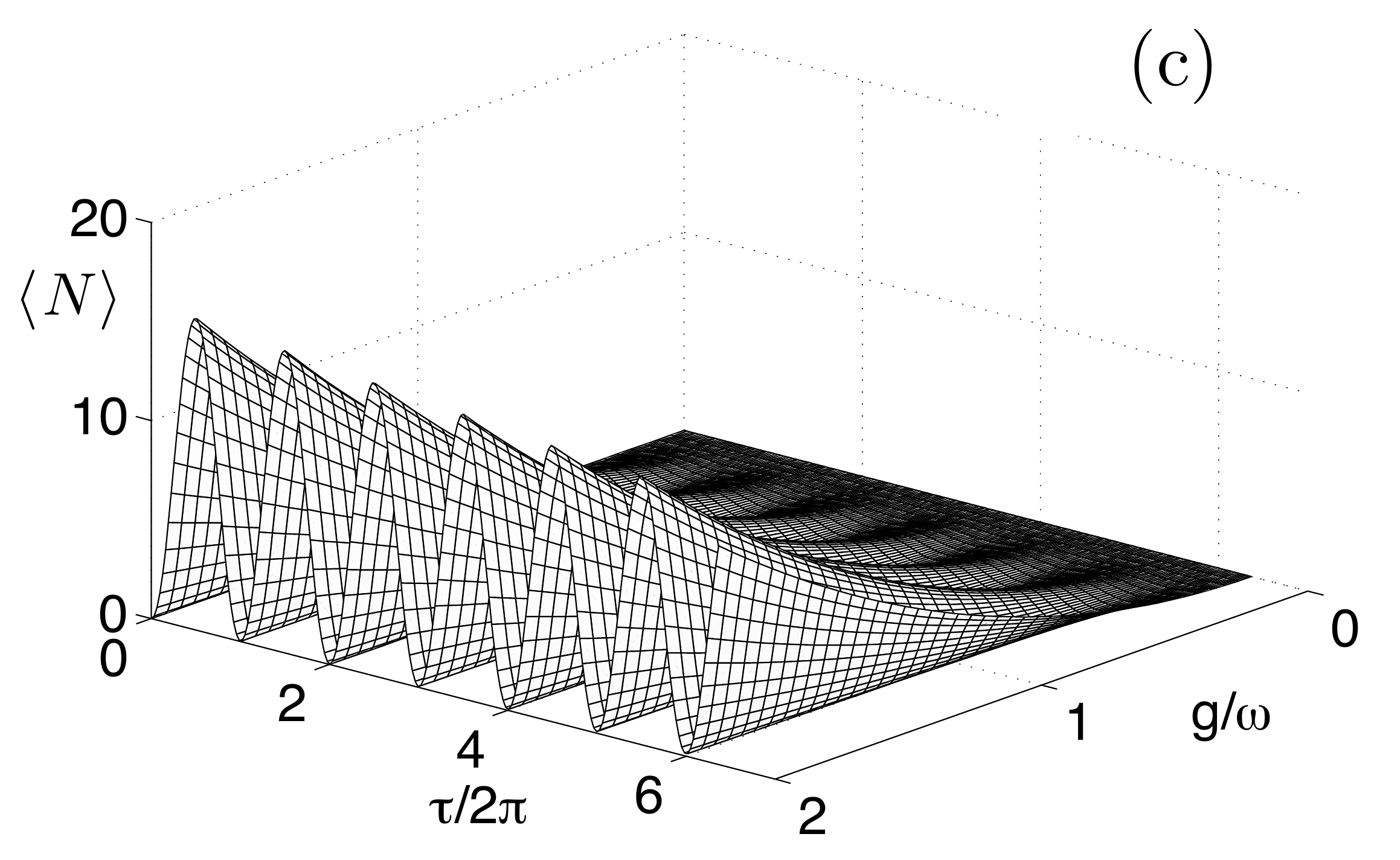}\\
\includegraphics[width=0.35\textwidth]{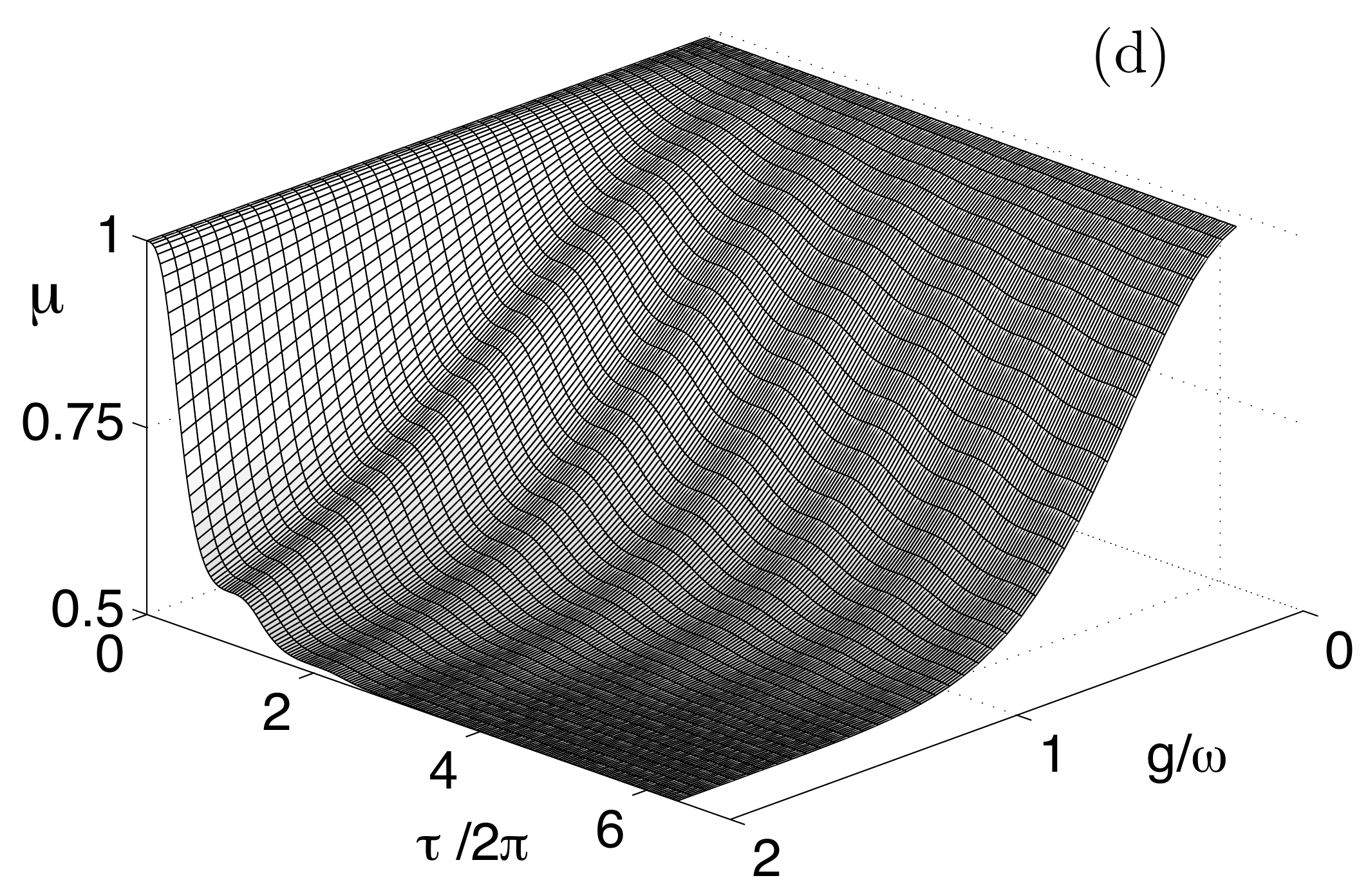}
\caption{\label{fig1}Effect of the parameter  $g/\omega$ on probabilities $P^{(+)}_{0}(t)$ (a), $P^{(-)}_{0}(t)$ (b) bosonic mode mean photon number $\langle N(t) \rangle$ (c), and purity $\mu(t)$ (d). We consider the off-resonance case $\Delta/\omega\simeq1$ and a dissipative decay rate $\kappa/\omega=0.01$.}
\end{figure}

Given the nature of the interaction containing both rotating and counter-rotating terms, the Hilbert space of the system is split in two parity chains~\cite{Casanova10}, associated with the eigenvalues $p=\pm 1$ of the parity operator $\hat{\Pi}\equiv-\hat{\sigma}_z(-1)^{\hat{a}^\dag\hat{a}}$, that are unconnected for $\kappa=0$. For instance, the states $\{\ket{g,2N},\ket{e,2N+1}\}$ have parity $p=1$, while the states $\{\ket{e,2N},\ket{g,2N+1}\}$ possess the opposite parity $p=-1$, where $N$ is an integer number. The probabilities that the system is in a state of one of the two
chains are
\begin{subequations}\begin{align}\label{PhotStat}
P^{(+)}_{n}(t)&=\frac{|\beta(t)|^{2n}}{n!}P_{g,0}(t),\\
P^{(-)}_{n}(t)&=\frac{|\beta(t)|^{2n}}{n!}P_{e,0}(t).
\end{align}\end{subequations}
We remark that in the unitary limit, $\kappa=0$, the decoherence function becomes $F(t)={\rm e}^{-2|\beta(t)|^2}$ so that we obtain $P_{g,0}(t)={\rm e}^{-|\beta(t)|^2}$ and $P_{e,0}(t)=0$. Hence, starting from $\ket{g,0}$, the evolved state vector remains in the subspace corresponding to parity chain $p=1$. Actually, starting from any superposition state of the system, the time evolution of $P^{(\pm)}_{n}(t)$ takes place independently in each parity
chain~\cite{Casanova10}. Here, for $\kappa > 0$, we see that $P_{e,0}(t)\neq 0$ and both parity chains are connected by the dissipative mechanism. It is straightforward to obtain analogous results starting from state $\ket{e,0}$, just replacing $F(t)$ by $-F(t)$.

\section{Analytical results}
\begin{figure}
\includegraphics[width=0.45\textwidth]{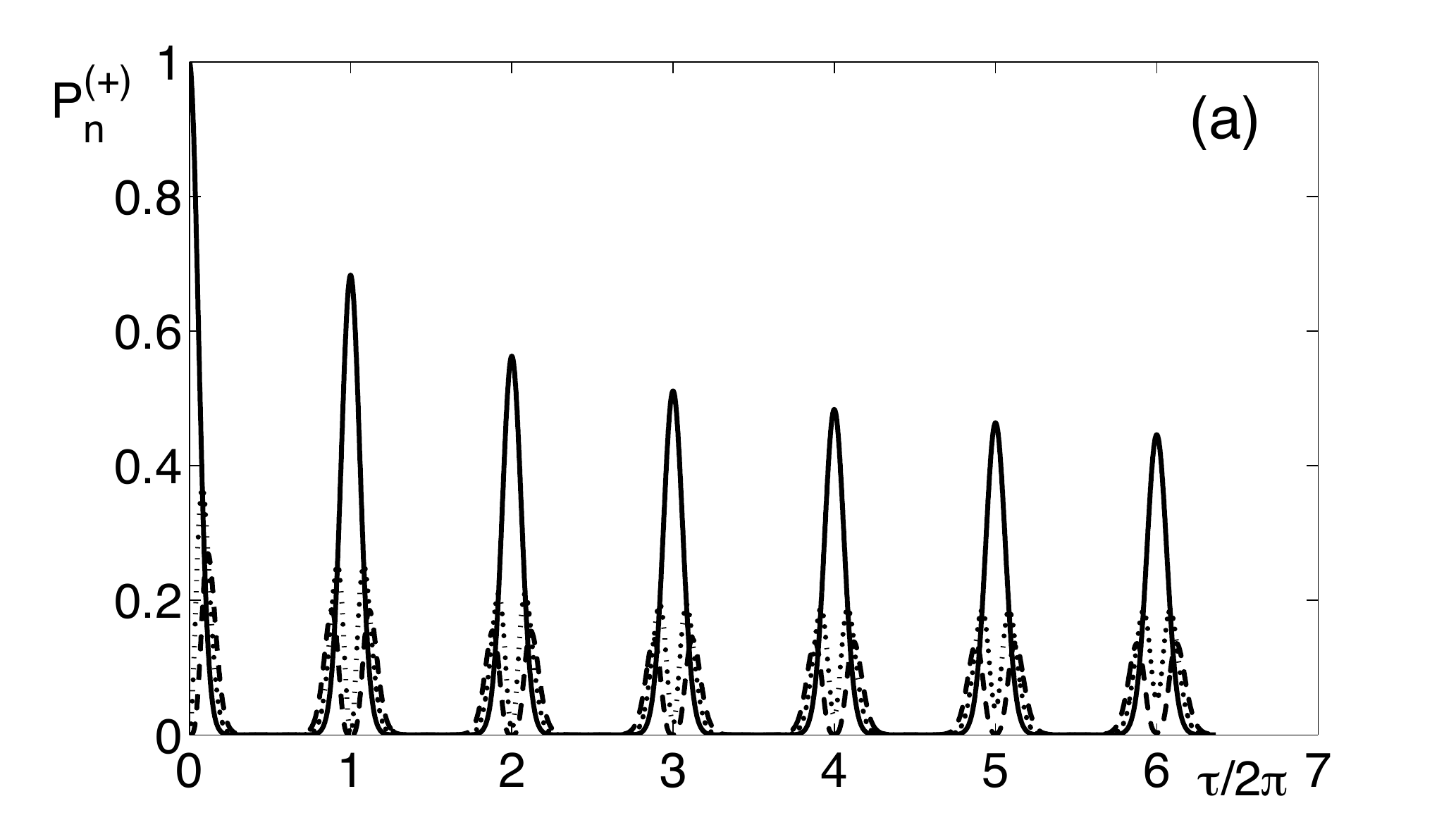}
\includegraphics[width=0.45\textwidth]{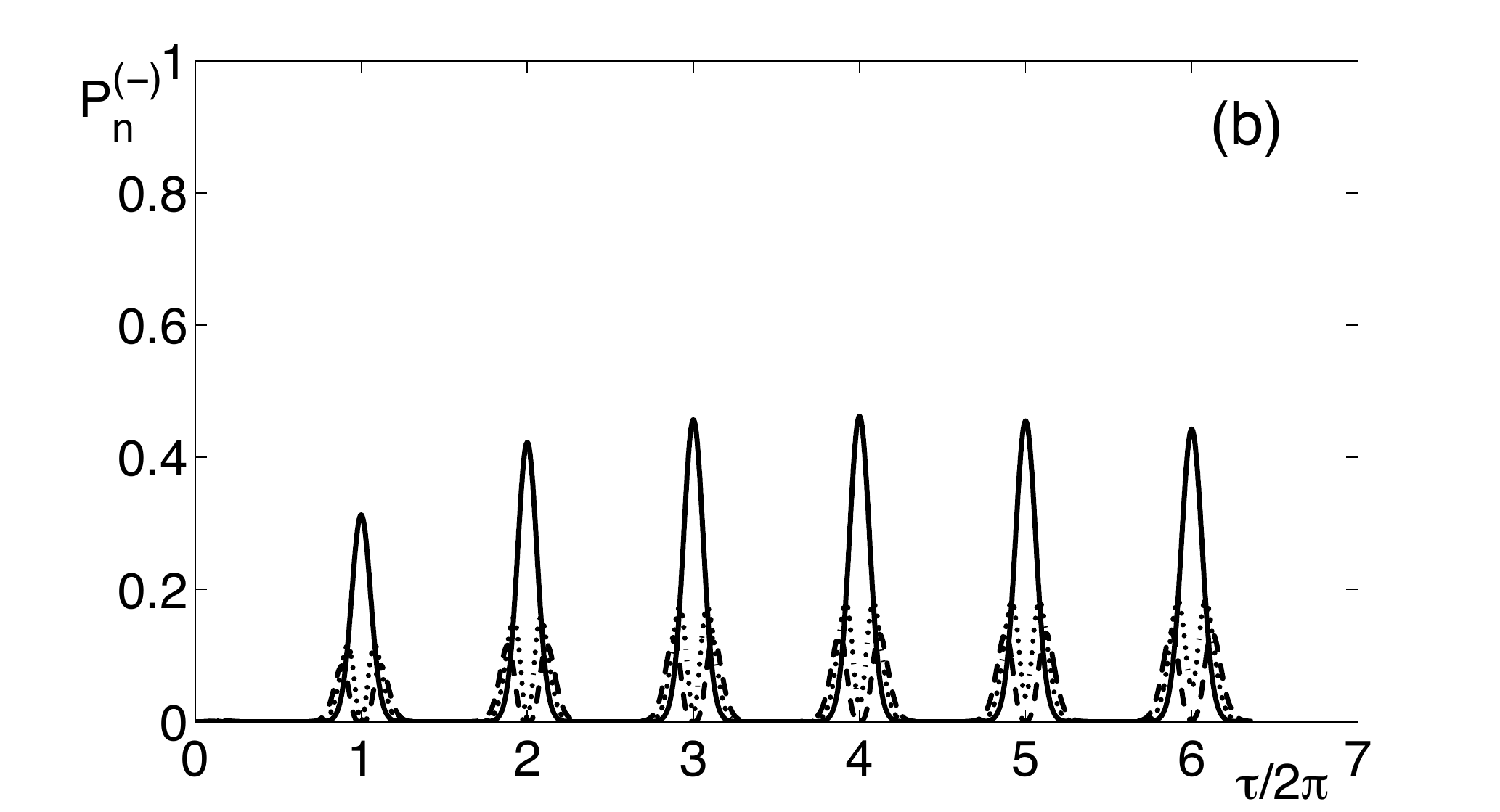}
\includegraphics[width=0.45\textwidth]{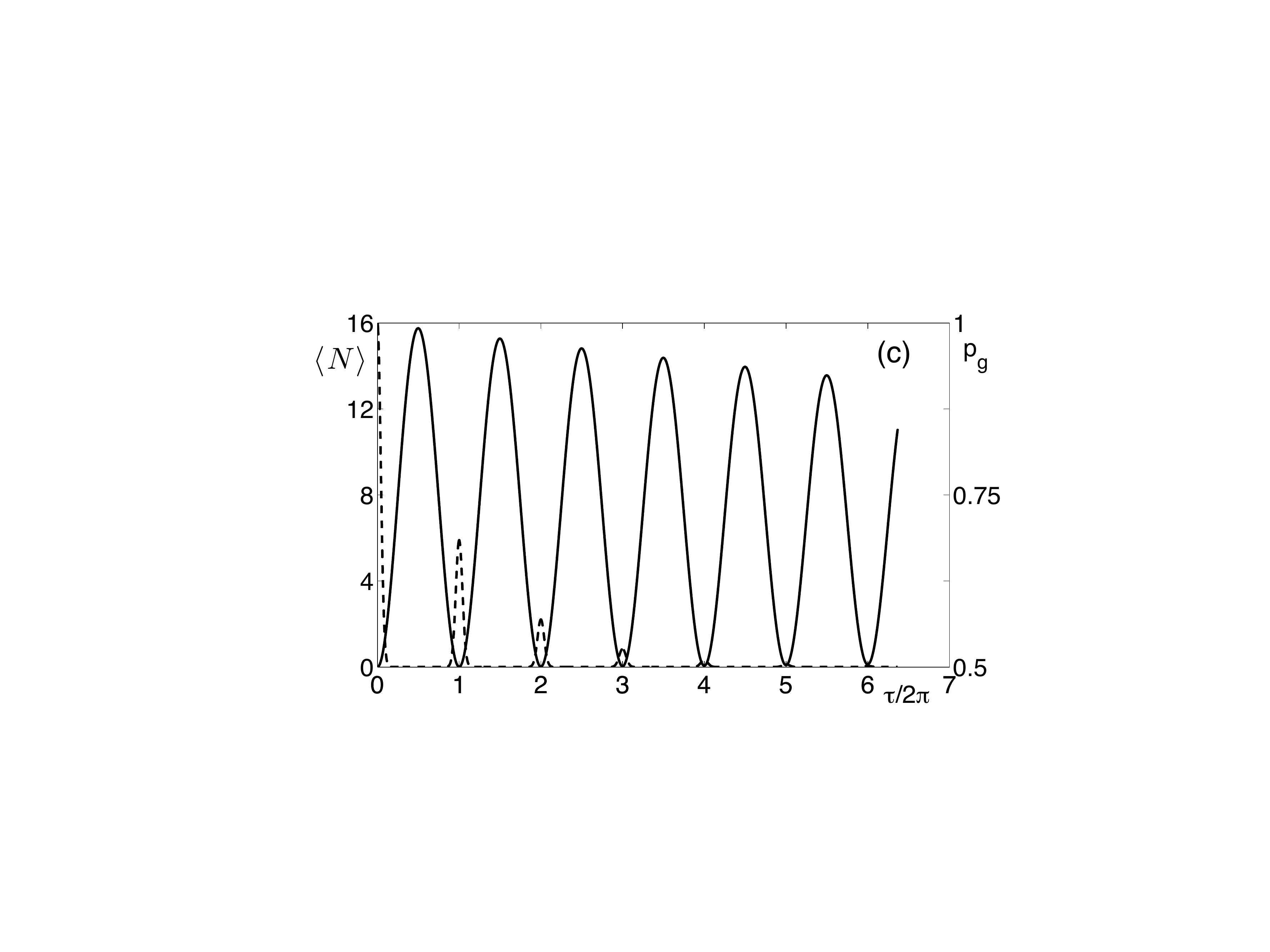}
\caption{\label{fig2} Analytic plot of time evolution of $P^{(+)}_{n}$ (a) and $P^{(-)}_{n}$ (b) for $n=0$ (solid), 1 (dotted), 2 (dashed). (c) behaviour of the subsystems: mean photon number $\langle N(t) \rangle$ (solid), probability of qubit lower state $P_g(t)$ (dashed). For the parameters $g/\omega=2$, $\Delta/\omega\simeq 1$
and $\kappa/\omega=0.01$. }
\end{figure}
In this section, we discuss the dynamics in the limit of large detuning $\Delta/\omega\simeq 1$ (i.e. $\omega_0/\omega\simeq 0$) as described by our analytical model. Let us recall that this model is only valid for dimensionless time $\tau=\omega t \ll 1/(1-\Delta/\omega)$. The focus of this discussion will be on the peculiar dynamics of the DSC regime~\cite{Casanova10}, such as collapses, revivals, and parity chains and how they are modified by the presence of dissipation, $\kappa$. 

Figure~\ref{fig1} shows how the DSC dynamics emerges under the condition $\Delta/\omega\simeq 1$, as one increases $g/\omega$. The first signatures are the appearance of collapses and revivals in the probability $P^{(+)}_{0}(t)$, and a periodic dynamics of the average photon number $\langle N(t) \rangle$, as it was already seen in Ref.~\cite{Casanova10}. The novel feature consists on the transfer of probability from the positive to the negative parity chain, evidenced in Fig.~\ref{fig1}b. This effect is induced by the presence of the small dissipation, $\kappa=0.01\omega$. Another consequence is the fact that the purity $\mu(t)$ decreases by increasing $g/\omega$ and that, in the DSC regime,
the system reaches quite rapidly a maximally mixed state. 
\begin{figure}
\includegraphics[width=0.49\linewidth]{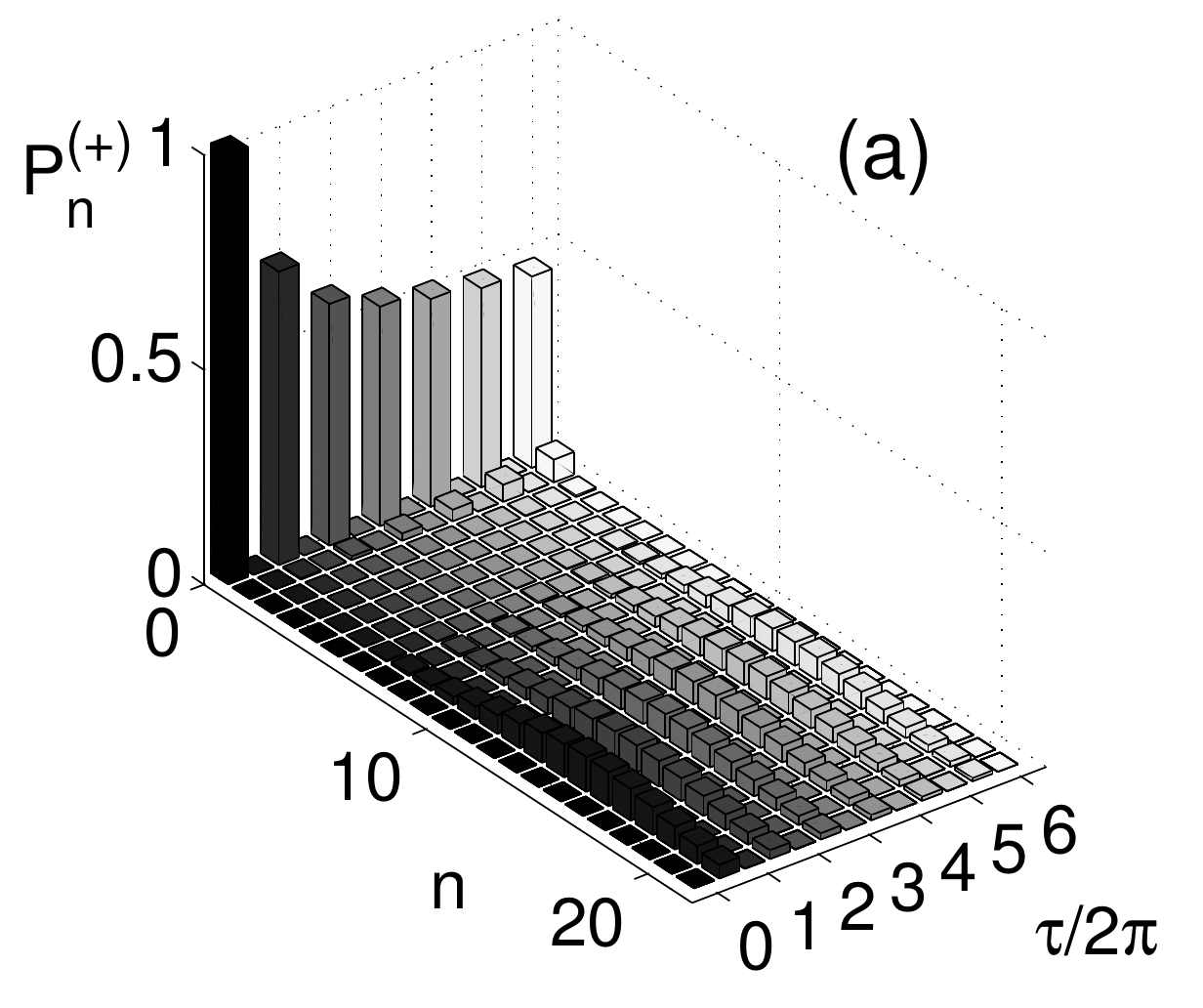}
\includegraphics[width=0.49\linewidth]{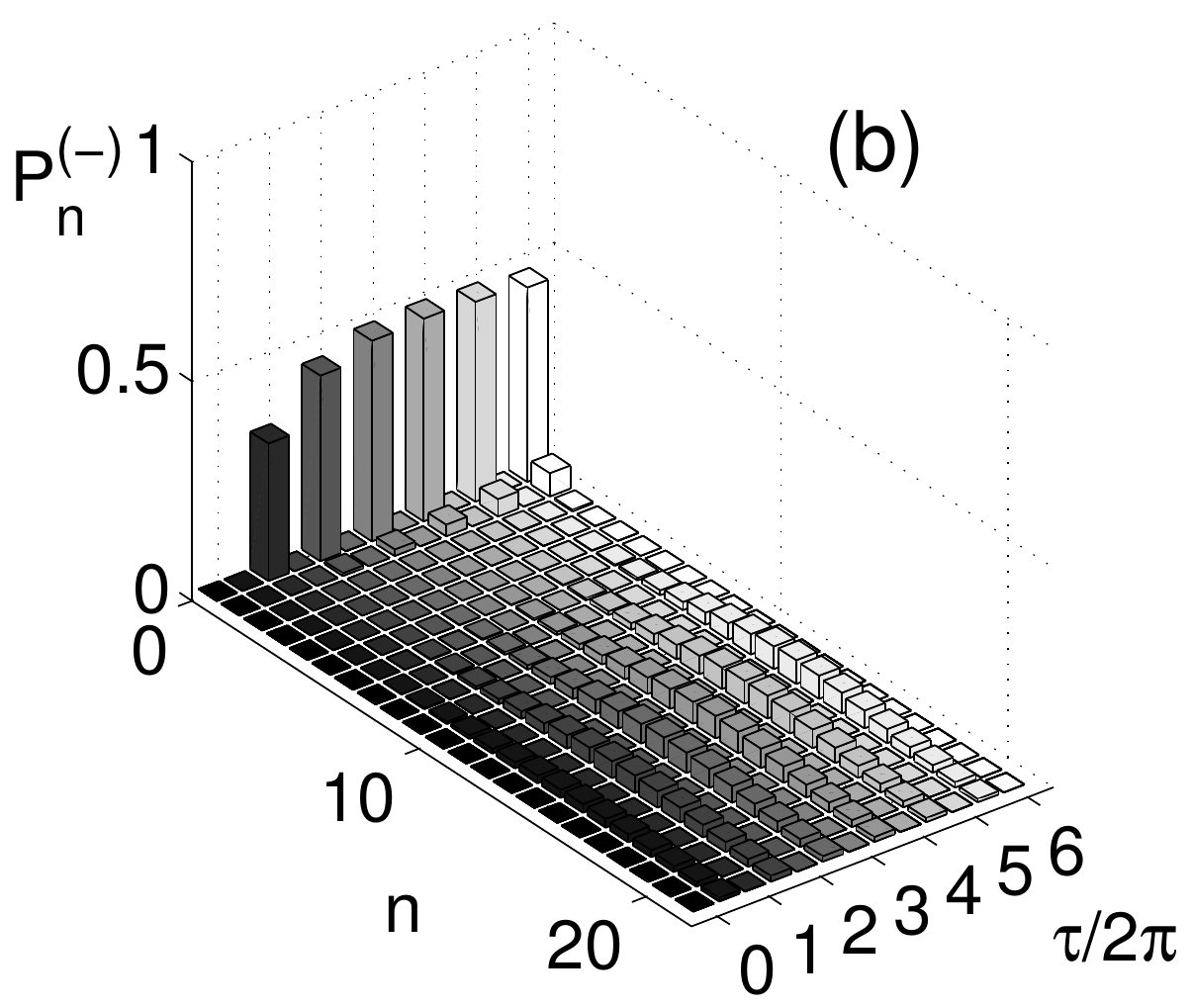}
\caption{\label{fig3} Analytic plot of probabilities $P^{(\pm)}_{n}(t)$ for $n\leq 20$ sampled at times $\tau_l=\pi l$ (with $l=0,\ldots,12$), with the same parameters as in Fig. \ref{fig2}.}
\end{figure}

\begin{figure}
\centering
\includegraphics[width=0.45\textwidth]{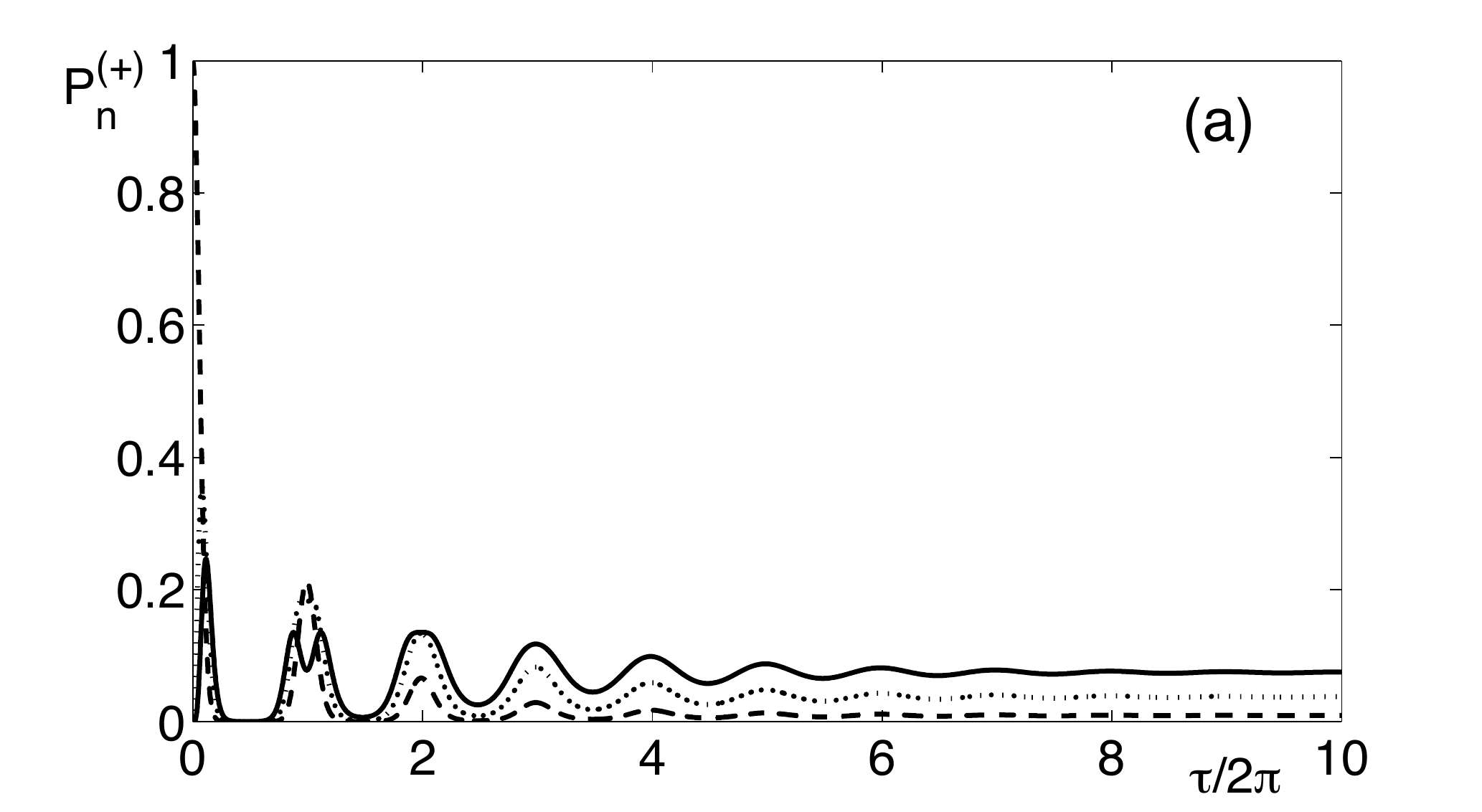}
\includegraphics[width=0.45\textwidth]{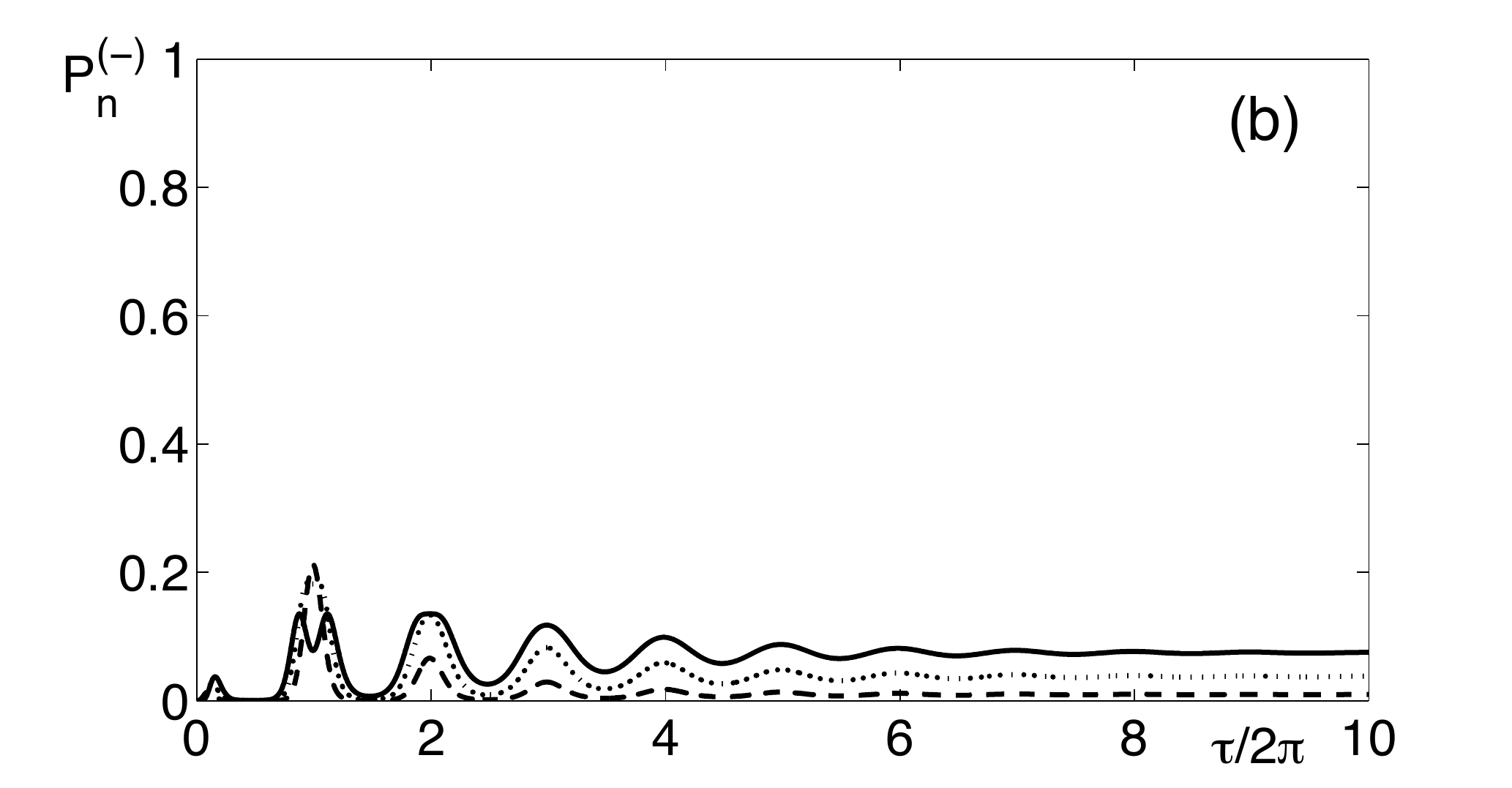}\\
\includegraphics[width=0.45\textwidth]{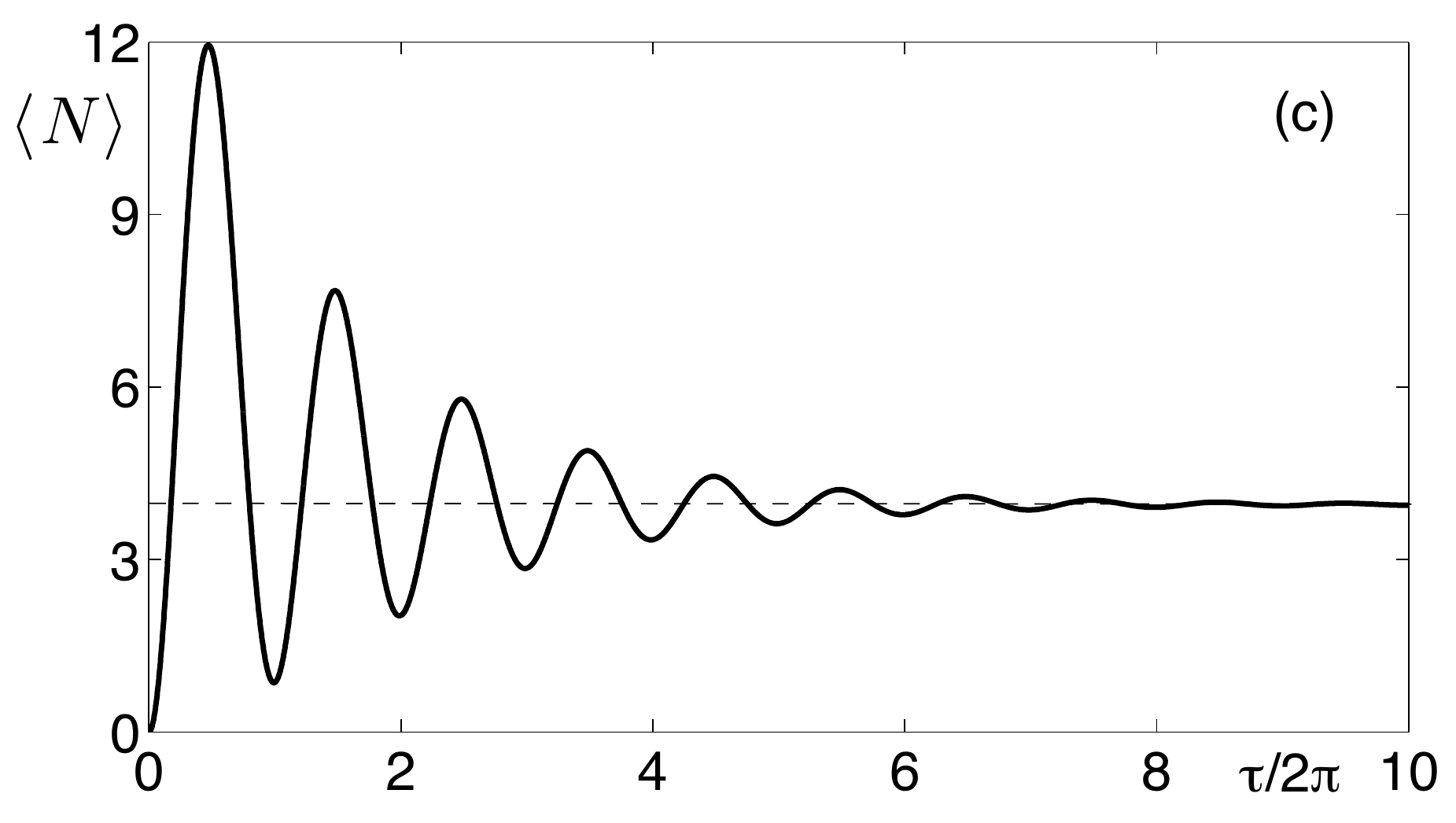}
\caption{\label{fig4}Analytic plots of the approach to the steady state of $P^{(+)}_{n}(t)$ (a) and $P^{(-)}_{n}(t)$ (b) for $n=0$ (dashed), 1 (dotted), 2 (solid), (c) Mean photon number $\langle N(t) \rangle$, with $g/\omega=2$, $\Delta/\omega\simeq 1$, and $\kappa/\omega=0.2$.}
\end{figure}

\begin{figure}
\includegraphics[width=0.45\textwidth]{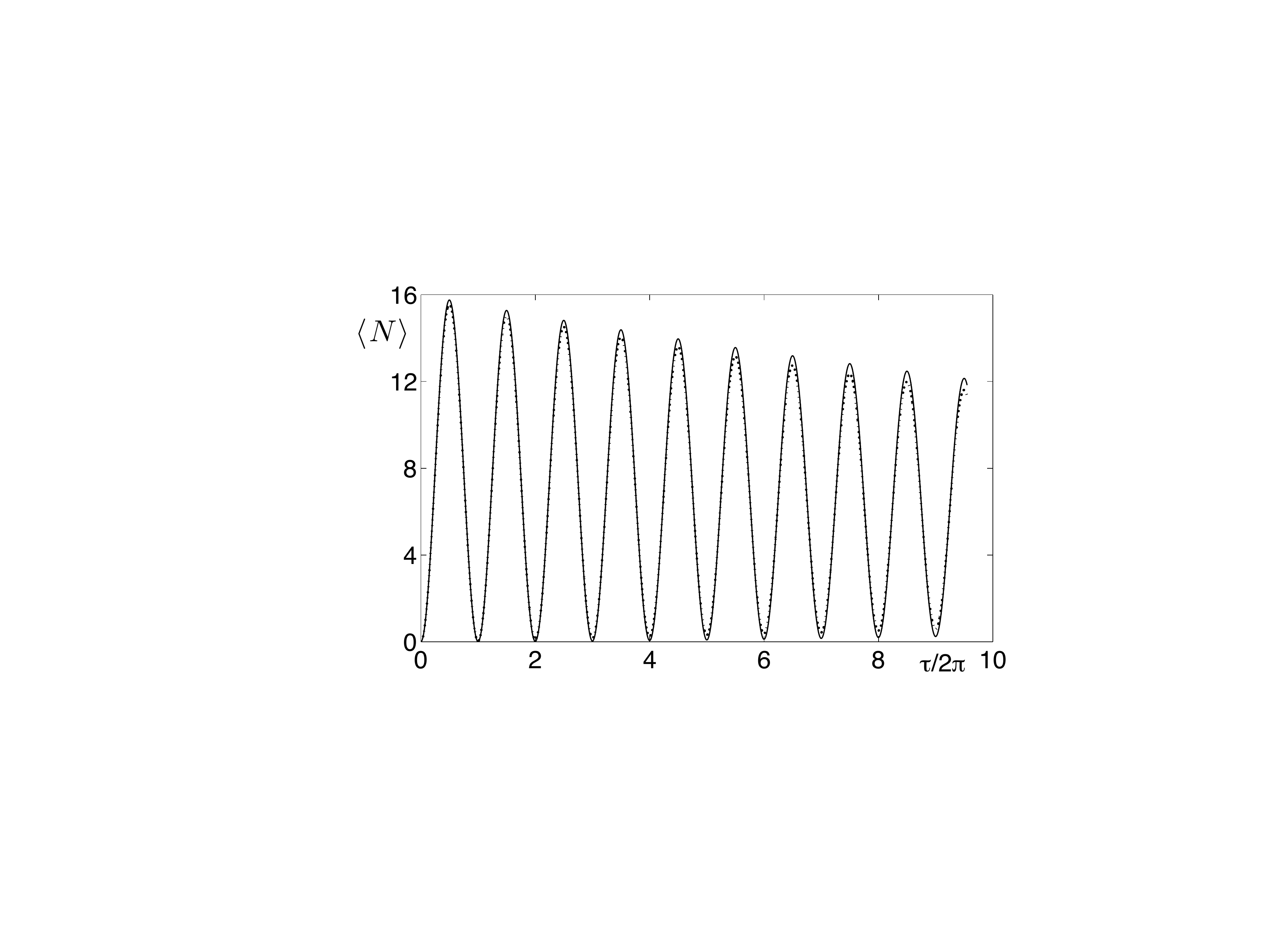}
\caption{\label{fig5} Comparison between the mean photon number $\langle N(t)\rangle$ from the analytical solution, in the limit $\Delta/\omega\simeq1$ (solid), and through MCWF method for $\Delta/\omega=0.8$ (dashed), in the DSC regime $g/\omega =2$ and
for $\kappa/\omega=0.01$.}
\end{figure}

\begin{figure}
\includegraphics[width=0.45\textwidth]{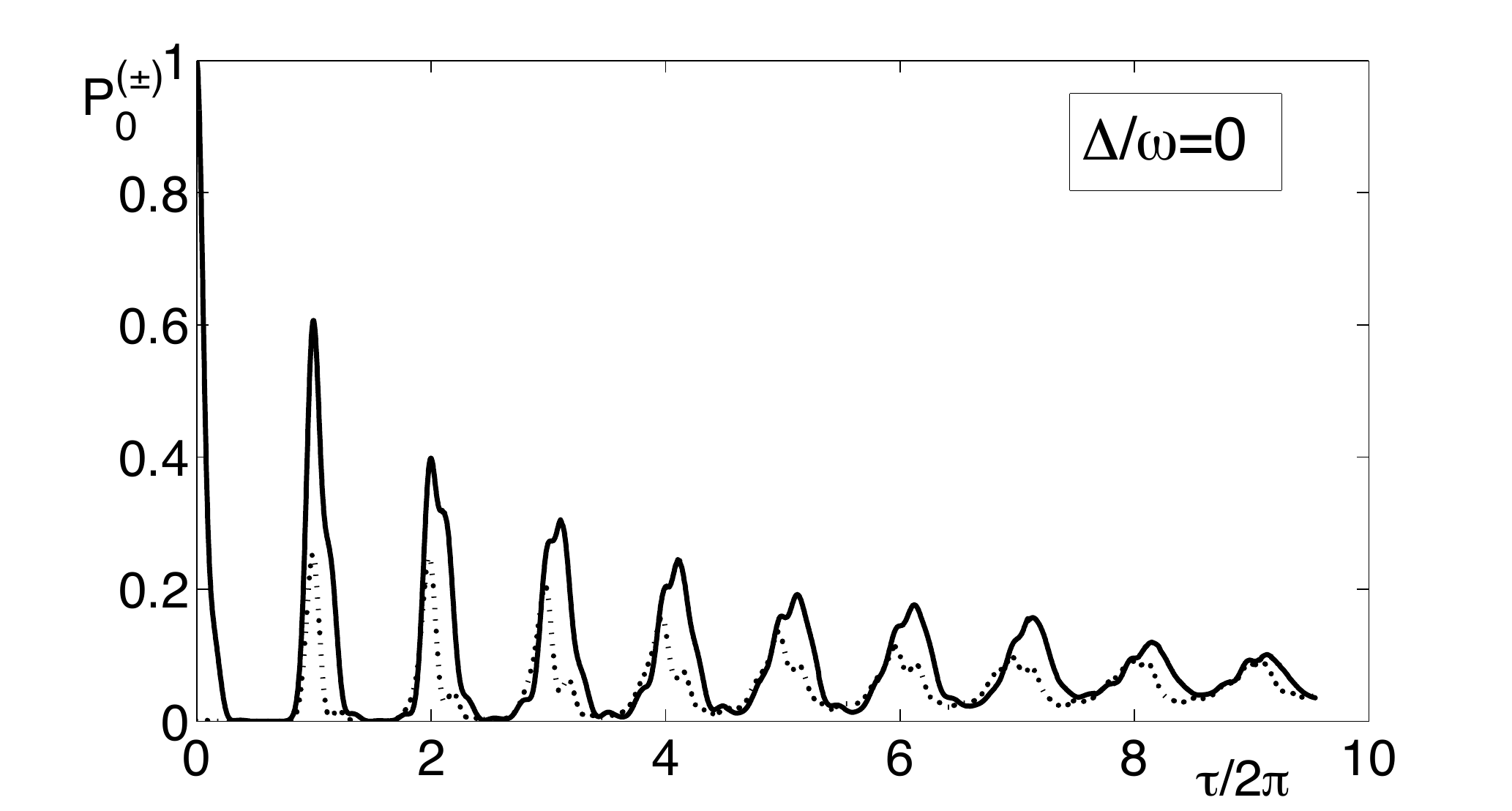}
\includegraphics[width=0.45\textwidth]{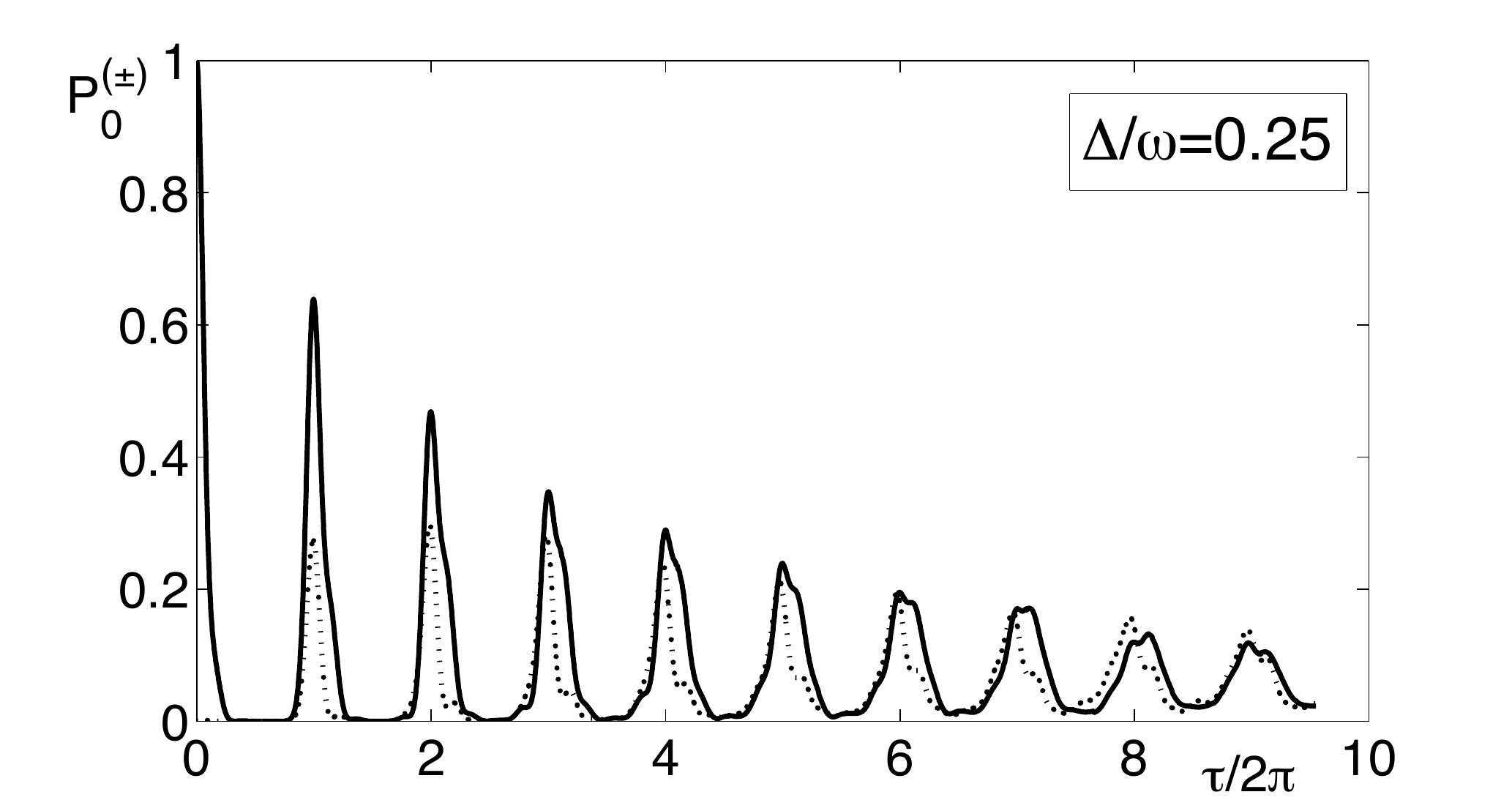}\\
\includegraphics[width=0.45\textwidth]{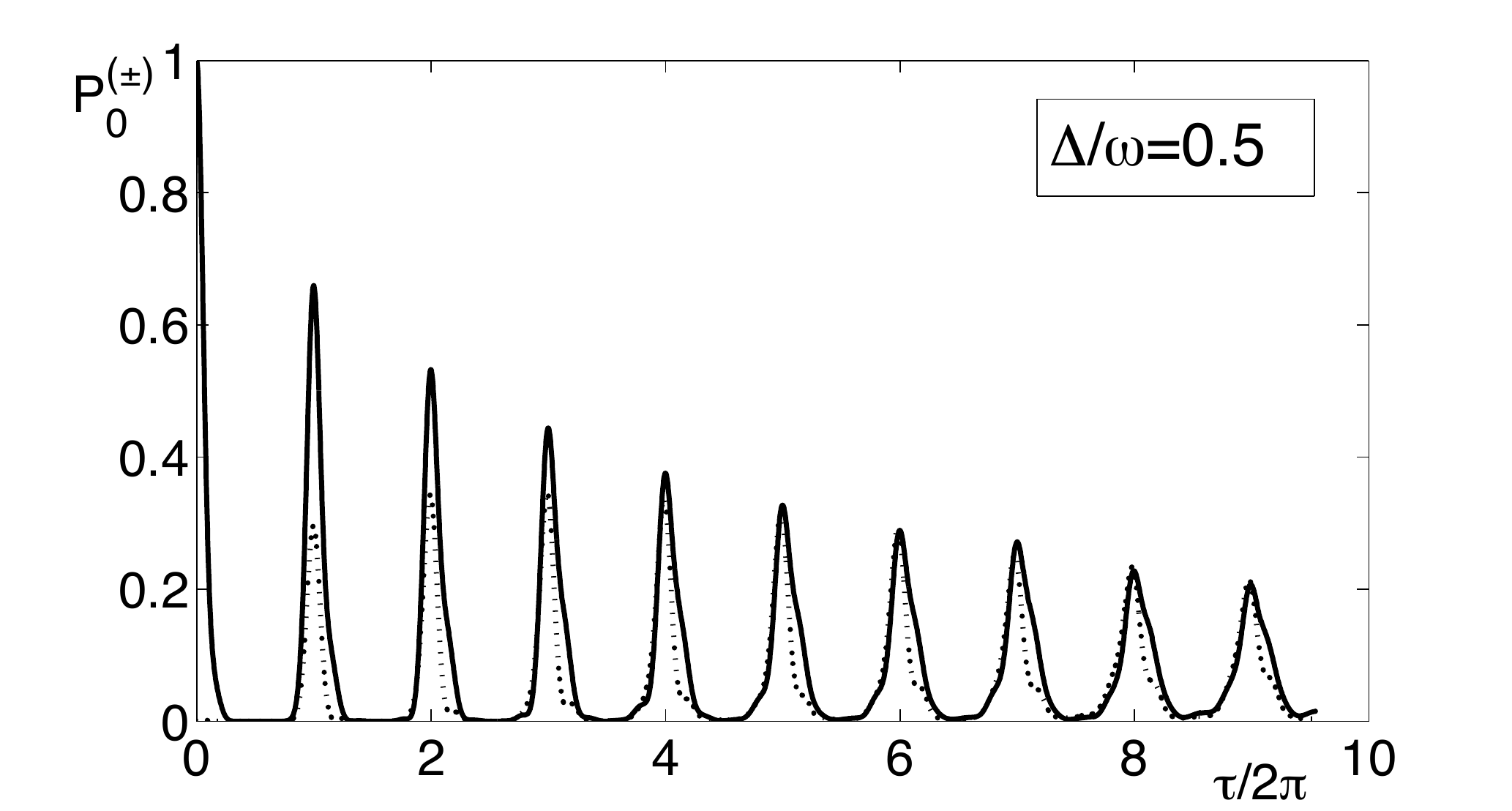}
\includegraphics[width=0.45\textwidth]{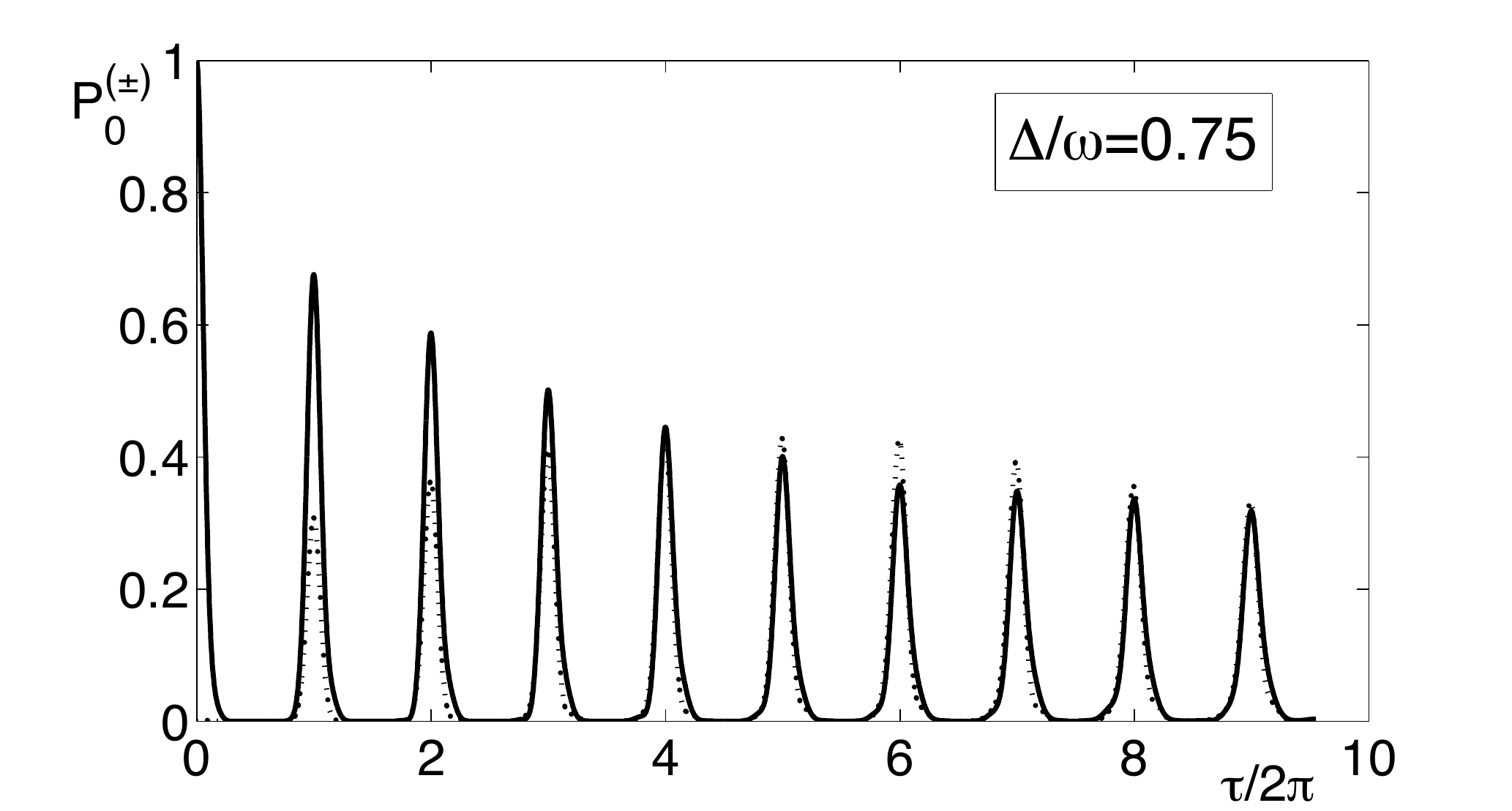}
\caption{\label{fig6} Probabilities $P^{(+)}_{0}(t)$ (solid) and $P^{(-)}_{0}(t)$ (dotted) as a function of $\Delta/\omega$ in the DSC regime, $g/\omega=2$ and $\kappa/\omega=0.01$.}
\end{figure}

In Fig.~\ref{fig2}, we show with greater detail the time evolution of the two parity chains, setting $g=2\omega$. Probabilities $P_n^{(+)}$ and $P_n^{(-)}$ exhibit revivals at $\tau=2\pi\times\mathbb{Z}^+$, and there is a progressive transfer of it from the positive to the negative parity chain. The consequence is that, at the end, the revivals of both chains exhibit the same structure and height, and that the qubit ends up with the same probability in the state $|g\rangle$ or $|e\rangle$ (cf. Fig.~\ref{fig2}c). 

In order to better illustrate the role of dissipation in connecting the two parity chains, we plot in Fig.~\ref{fig3} different time shots of the statistics $P^{(\pm)}_{n}(t)$ for large enough values of $n$. We notice that, starting from $\ket{g,0}$, the two chains are connected via the dissipative channel resulting in an incoherent mixture of them, breaking their independent dynamics stemming from the unitary case.

In our analytical framework, we can compute the steady state of the system,
\begin{equation}\label{steadystate}
\hat{\rho}_{S}=\frac{1}{2}\Big[\ketbra{\beta_{S}}{\beta_{S}}\otimes\ketbra{+}{+}+\ketbra{-\beta_{S}}{-\beta_{S}}\otimes\ketbra{-}{-} \Big ],
\end{equation}
where $\beta_{S}= - \imm g / z $ is the steady amplitude of the field coherent state. The structure of the steady state is
noteworthy, since it is a statistical mixture of two parts, one that associates $\ket{\beta_{S}}$ to the qubit state $\ket{+}$ and the other connects
$\ket{-\beta_{S}}$ to $\ket{-}$.

\begin{figure}
\includegraphics[width=0.45\textwidth]{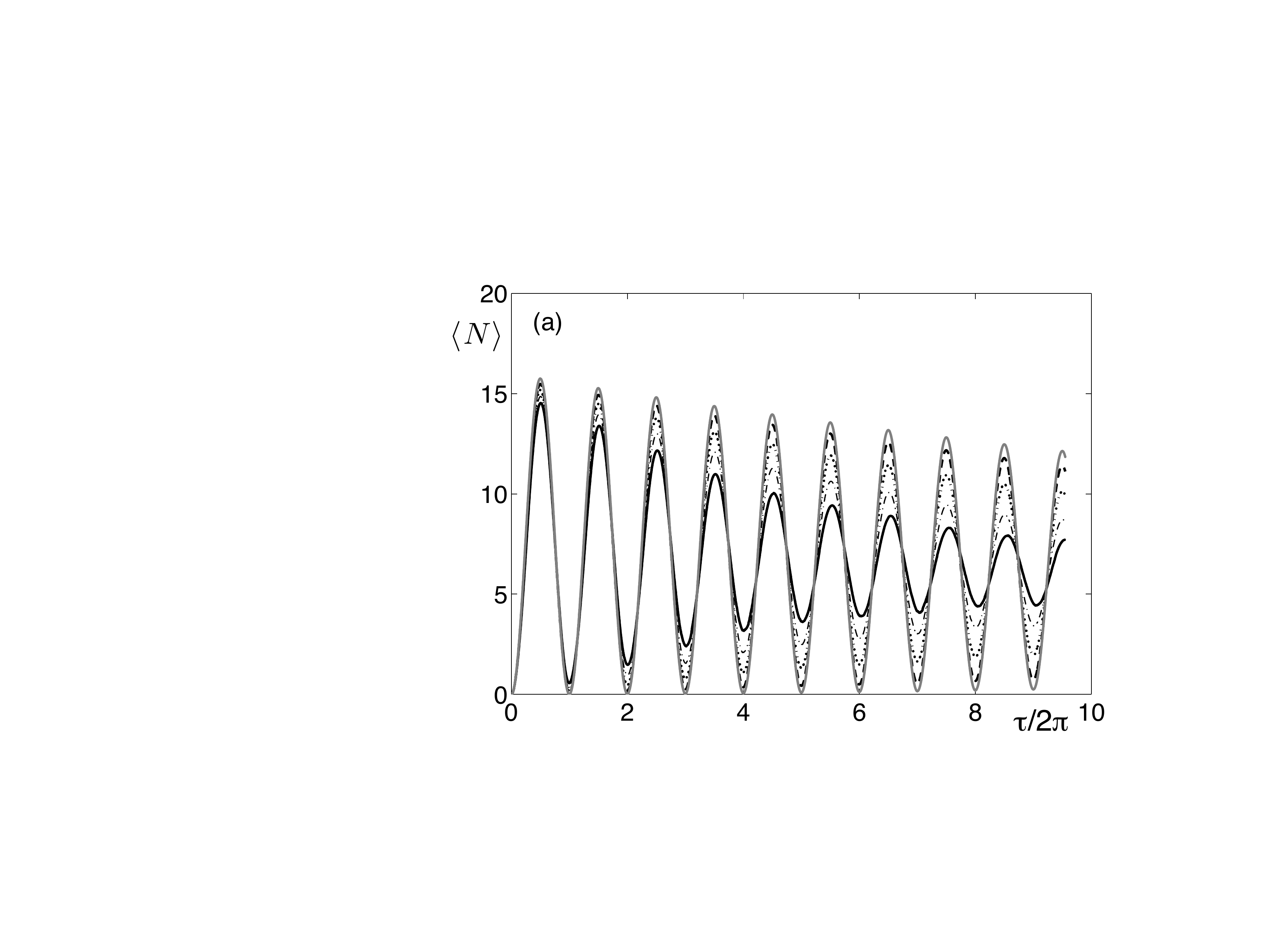}
\caption{\label{fig7} Effect of detuning in the DSC regime $g/\omega=2$ and $\kappa/\omega=0.01$. Plots of $\langle N(t) \rangle$: analytical (solid-gray) and numerical solutions for $\Delta/\omega=0.75$ (dashed), $\Delta/\omega=0.5$ (dotted), $\Delta/\omega=0.25$ (dash-dotted), $\Delta/\omega=0$ (solid-black).}
\end{figure}

\begin{figure}
\includegraphics[width=0.45\textwidth]{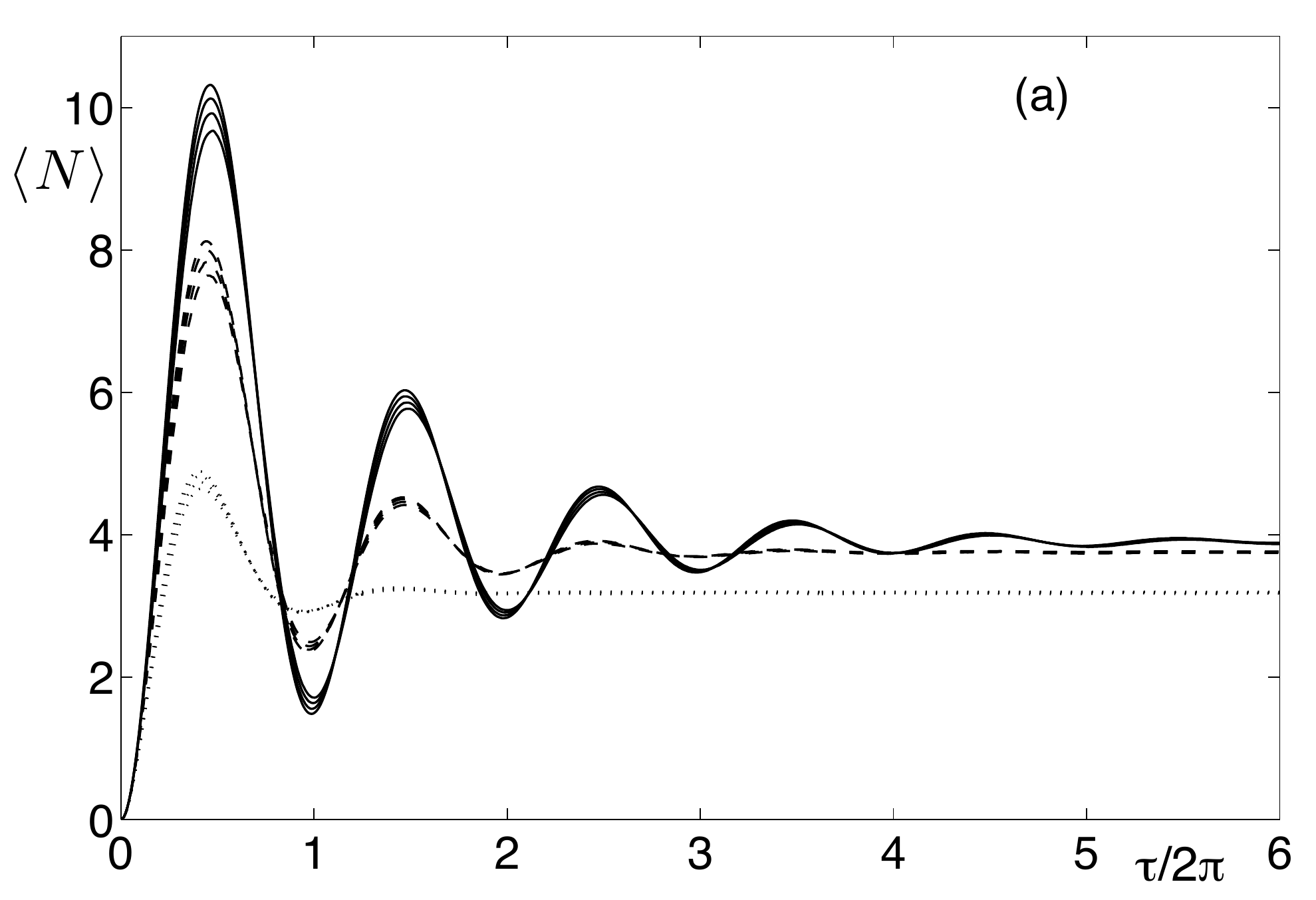}
\includegraphics[width=0.45\textwidth]{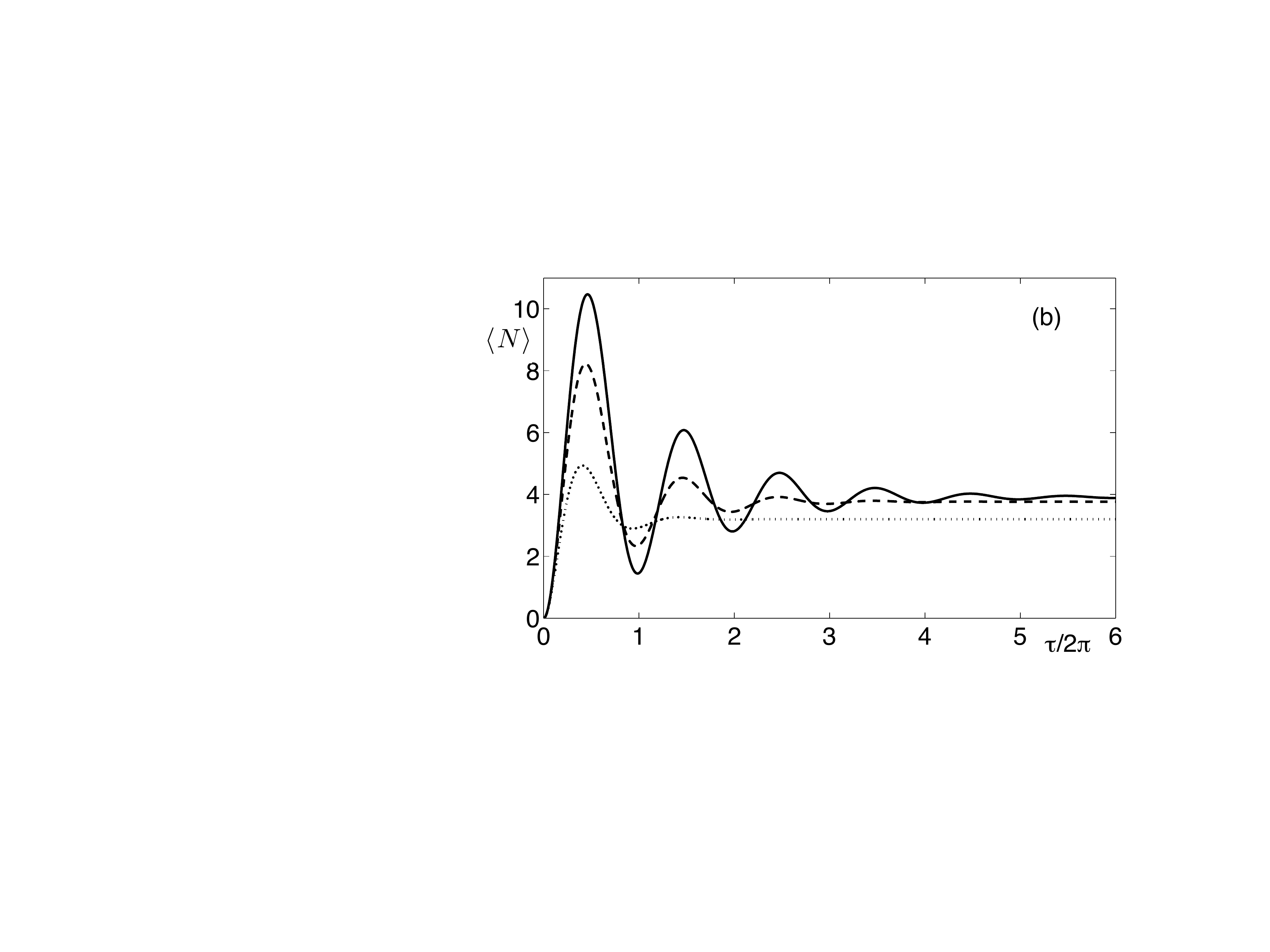}
\caption{\label{fig8} Mean photon number $\langle N(t) \rangle$ for different values of parameter $\kappa/\omega$: 0.3 (solid), 0.5(dashed), 1(dotted) and for the coupling parameter $g/\omega=2$. (a) Numerical results composed by four curves for each $\kappa/\omega$ value, corresponding, from top to bottom, to $\Delta/\omega=0.75, 0.5, 0.25, 0$, (b) Analytical solution in the far off resonant limit $\Delta/\omega\cong 1$.}
\end{figure}

This means that the system relaxes upon the two states $\ket{\pm,\pm\beta_S}$, which have the same energy mean value $\lambda_{\rho_S}=\hbar\Delta|\beta_S|^2+\hbar g (\beta_S+\beta_S^*)$. In Fig.~\ref{fig4}, we show how the first elements of the two parity chains $P^{(\pm)}_{n}(t)$ ($n=0,1,2$) and the mean photon number $\langle N(t) \rangle$ approach the steady state for relatively large values of decay rate $\kappa/\omega$. We remark that $P^{(\pm)}_{0}(t)$ vanish at long times, while the other probabilities reach a non-zero constant value, so that the steady state Poissonian photon statistics $P_n(t)=P^{(+)}_{n}(t)+P^{(-)}_{n}(t)$ has a mean value $|\beta_S|^2=4g^2/(\kappa^2+4\Delta^2)$,
with $\Delta/\omega\simeq 1$.

\section{Numerical results for $0\leq\Delta/\omega<1$}
The role of this section is to investigate numerically all regimes of the detuning $\Delta$ using the Monte Carlo wavefunction method~\cite{MCWF}. As a first example, we shall discuss a set of simulations for small values of the detuning $\Delta/\omega=0.8$, for which the analytical solution may still be valid at short times $\tau\ll 5$. These are values that may be achieved for experimental circuit QED setups~\cite{Niemczyk10,Pol10}, selecting a resonator of $\omega\sim 6$~GHz and $\omega_0=1.2$~GHz. The results of the simulation are shown in Fig.~\ref{fig5} where we plot the Monte Carlo simulation together with the analytical ansatz. It is evident that both solutions are very close, with an error $< 4\%$ for $\tau\le 9$, justifying the approximations of Sec.~II, at least for $0.8\le\Delta/\omega\le 1$.  

We investigate now the effect of moving towards the full resonance $\Delta/\omega=0$. The behavior of parity chain probabilities $P^{(\pm)}_{n}(t)$ is heavily affected by the reduction of
the detuning values $\Delta/\omega$. As an example, in Fig.~\ref{fig6}, we show that the probabilities $P^{(\pm)}_{0}(t)$ are progressively spoiled, losing the symmetry of the ideal case $\Delta/\omega\simeq 1$ (see Fig.~\ref{fig3}a).  We outline that the probabilities $P^{(\pm)}_{n}(t)$ with $n>0$ present irregular oscillations that do not vanish for long times. Despite the distortions of $P^{(\pm)}_{n}(t)$, we show in Fig.~\ref{fig7} that the mean photon number $\langle N(t) \rangle$ exhibits regular oscillations. The only effect of decreasing the rate $\Delta/\omega$, towards a resonant condition, is to reduce the amplitude of oscillations and to induce a small shift in the peak times. In order to quantify this effect, we estimate the error in each curve at $\tau=8.5$ with respect to the analytical solution. The percentage of relative differences are $7\%$ ($\Delta/\omega=0.75$), $17\%$ ($0.5$), $29\%$ ($0.25$) and $38\%$ ($0$). These features can be explained as an internal dephasing mechanism due to the qubit free energy~\cite{Casanova10}.

Finally, let us consider the simultaneous effect of different decay rates and detunings. Figure~\ref{fig8} shows the results of the numerical simulation for the mean photon number. We notice that the effect of a decreasing detuning becomes almost negligible in the presence of dissipation. Moreover, the asymptotic value of $\langle N(t)\rangle$ agree very well with the analytical predictions of Sec.~II.

\section{Conclusions}
We have studied the DSC regime of the quantum Rabi model with a zero-temperature bath interacting with the single quantized mode. We have found analytical solutions for the off-resonant case, describing the dissipative dynamics that induces incoherent mixtures of both parity chains. Furthermore, we have presented a numerical analysis for the near-resonant case where analytical solutions are not available. Finally, we have estimated the limits of our analytical solutions as $\Delta/\omega$ decreases, among other features. 
\newline
\newline
\section*{Acknowledgments}
G.R. acknowleges funding from Juan de la Cierva MICINN Program, J. C. from Basque Government BFI08.211, J.J.G.-R. from Spanish projects MICINN FIS2009-10061 and QUITEMAD, and E.S. from Basque Government IT472-10, Spanish MICINN FIS2009-12773-C02-01, SOLID and CCQED European projects. The authors deeply regret the loss of their colleague and coauthor Federico Casagrande who sadly passed away before the submission of this paper.

\end{document}